\documentclass[aps,prr,a4paper,superscriptaddress,twocolumn,showpacs,amsmath,amssymb,longbibliography]{revtex4-1}

\usepackage{bm}

\usepackage{graphicx,epsfig}
\usepackage[usenames]{color}

\usepackage[english]{babel}

\usepackage{array}

\usepackage{dsfont}
\usepackage{multirow}

\usepackage{csquotes}
\MakeOuterQuote{"}

\usepackage{aliascnt}

\allowdisplaybreaks

\begin{document}

\newcommand{\btheta}{\bm{\theta}}
\newcommand{\bdtheta}{\mathbf{d}\bm{\theta}}

\newcommand{\E}{\mathcal{E}}
\newcommand{\G}{\mathcal{G}}
\newcommand{\Lag}{\mathcal{L}}
\newcommand{\M}{\mathcal{M}}
\newcommand{\N}{\mathcal{N}}
\newcommand{\U}{\mathcal{U}}
\newcommand{\R}{\mathcal{R}}
\newcommand{\F}{\mathcal{F}}
\newcommand{\V}{\mathcal{V}}
\newcommand{\C}{\mathcal{C}}
\newcommand{\I}{\mathcal{I}}
\newcommand{\s}{\sigma}
\newcommand{\up}{\uparrow}
\newcommand{\dw}{\downarrow}
\newcommand{\h}{\hat{\mathcal{H}}}
\newcommand{\himp}{\hat{h}}
\newcommand{\g}{\mathcal{G}^{-1}_0}
\newcommand{\D}{\mathcal{D}}
\newcommand{\A}{\mathcal{A}}
\newcommand{\projs}{\hat{\mathcal{S}}_d}
\newcommand{\proj}{\hat{\mathcal{P}}_d}
\newcommand{\K}{\textbf{k}}
\newcommand{\Q}{\textbf{q}}
\newcommand{\T}{\tau_{\ast}}
\newcommand{\io}{i\omega_n}
\newcommand{\eps}{\varepsilon}
\newcommand{\+}{\dag}
\newcommand{\su}{\uparrow}
\newcommand{\giu}{\downarrow}
\newcommand{\0}[1]{\textbf{#1}}
\newcommand{\ca}{c^{\phantom{\dagger}}}
\newcommand{\cc}{c^\dagger}

\newcommand{\fa}{f^{\phantom{\dagger}}}
\newcommand{\fc}{f^\dagger}

\newcommand{\aaa}{a^{\phantom{\dagger}}}
\newcommand{\aac}{a^\dagger}
\newcommand{\bba}{b^{\phantom{\dagger}}}
\newcommand{\bbc}{b^\dagger}
\newcommand{\da}{{d}^{\phantom{\dagger}}}
\newcommand{\dc}{{d}^\dagger}
\newcommand{\ha}{h^{\phantom{\dagger}}}
\newcommand{\hc}{h^\dagger}
\newcommand{\be}{\begin{equation}}
\newcommand{\ee}{\end{equation}}
\newcommand{\bea}{\begin{eqnarray}}
\newcommand{\eea}{\end{eqnarray}}
\newcommand{\ba}{\begin{eqnarray*}}
\newcommand{\ea}{\end{eqnarray*}}
\newcommand{\dagga}{{\phantom{\dagger}}}
\newcommand{\bR}{\mathbf{R}}
\newcommand{\bQ}{\mathbf{Q}}
\newcommand{\bq}{\mathbf{q}}
\newcommand{\bqp}{\mathbf{q'}}
\newcommand{\bk}{\mathbf{k}}
\newcommand{\bh}{\mathbf{h}}
\newcommand{\bkp}{\mathbf{k'}}
\newcommand{\bp}{\mathbf{p}}
\newcommand{\bL}{\mathbf{L}}
\newcommand{\bRp}{\mathbf{R'}}
\newcommand{\bx}{\mathbf{x}}
\newcommand{\by}{\mathbf{y}}
\newcommand{\bz}{\mathbf{z}}
\newcommand{\br}{\mathbf{r}}
\newcommand{\Ima}{{\Im m}}
\newcommand{\Rea}{{\Re e}}
\newcommand{\Pj}[2]{|#1\rangle\langle #2|}
\newcommand{\ket}[1]{\vert#1\rangle}
\newcommand{\bra}[1]{\langle#1\vert}
\newcommand{\setof}[1]{\left\{#1\right\}}
\newcommand{\fract}[2]{\frac{\displaystyle #1}{\displaystyle #2}}
\newcommand{\Av}[2]{\langle #1|\,#2\,|#1\rangle}
\newcommand{\av}[1]{\langle #1 \rangle}
\newcommand{\Mel}[3]{\langle #1|#2\,|#3\rangle}
\newcommand{\Avs}[1]{\langle \,#1\,\rangle_0}
\newcommand{\eqn}[1]{(\ref{#1})}
\newcommand{\Tr}{\mathrm{Tr}}

\newcommand{\Vb}{\bar{\mathcal{V}}}
\newcommand{\Vd}{\Delta\mathcal{V}}

\newcommand{\Rb}{\bar{R}}
\newcommand{\Rd}{\Delta R}
\newcommand{\ocrev}[1]{{\color{cyan}{#1}}}
\newcommand{\occom}[1]{{\color{red}{#1}}}

\title{Error mitigation in variational quantum eigensolvers using tailored probabilistic machine learning}

\author{Tao Jiang}
\affiliation{Ames National Laboratory, Ames, Iowa 50011, USA}

\author{John Rogers}
\affiliation{Department of Physics and Astronomy, Texas A\&M University, College Station, Texas 77845, USA}

\author{Marius S. Frank}
\affiliation{Department of Chemistry, Aarhus University, 8000, Aarhus C, Denmark}
\author{Ove Christiansen}
\affiliation{Department of Chemistry, Aarhus University, 8000, Aarhus C, Denmark}
\author{Yong-Xin Yao}
\altaffiliation{Corresponding author: ykent@iastate.edu}
\affiliation{Ames National Laboratory, Ames, Iowa 50011, USA}
\affiliation{Department of Physics and Astronomy, Iowa State University, Ames, Iowa 50011, USA}
\author{Nicola Lanat\`a}
\altaffiliation{Corresponding author: nxlsps@rit.edu}
\affiliation{School of Physics and Astronomy, Rochester Institute of Technology, 84 Lomb Memorial Drive, Rochester, New York 14623, USA}
\affiliation{Center for Computational Quantum Physics, Flatiron Institute, New York, New York 10010, USA}

\date{\today}

\begin{abstract}

Quantum computing technology has the potential to revolutionize the simulation of materials and molecules in the near future. A primary challenge in achieving near-term quantum advantage is effectively mitigating the noise effects inherent in current quantum processing units (QPUs). This challenge is also decisive in the context of quantum-classical hybrid schemes employing variational quantum eigensolvers (VQEs) that have attracted significant interest in recent years. In this work, we present a novel method that employs parametric Gaussian process regression (GPR) within an active learning framework to mitigate noise in quantum computations, focusing on VQEs. Our approach, grounded in probabilistic machine learning, exploits a custom prior based on the VQE ansatz to capture the underlying correlations between VQE outputs for different variational parameters, thereby enhancing both accuracy and efficiency. We demonstrate the effectiveness of our method on a 2-site Anderson impurity model and a 8-site Heisenberg model, using the IBM open-source quantum computing framework, Qiskit, showcasing substantial improvements in the accuracy of VQE outputs while reducing the number of direct QPU energy evaluations. This work contributes to the ongoing efforts in quantum error mitigation and optimization, bringing us a step closer to realizing the potential of quantum computing in quantum matter simulations.

\end{abstract}

\maketitle

\section{Introduction}

Quantum computers hold the promise of dramatically enhancing our ability to simulate quantum-mechanical systems. Computational chemistry is expected to be one of the most promising fields to benefit significantly from quantum technologies within the next few years~\cite{preskill2018quantum, aspuru2005simulated, QC-molecules1,QC-molecule2,QC-molecule3,QC-molecule4}, as the number of qubits required to represent the active degrees of freedom for small molecules is relatively low compared to that of larger systems. Furthermore, quantum embedding (QE) methods~\cite{Kotliar-Science,quantum-embedding-review} could allow us to benefit from quantum devices for simulating larger systems (including molecules and periodic materials) by employing QPUs to handle only the most important degrees of freedom, while treating others (at the mean-field level) on classical devices~\cite{qe-ntqc1,HybridQC,YY-qc,jaderberg2020minimum}.

A major obstacle to realizing this ambitious program is the insufficient reliability of data produced by quantum devices, even though several effective error-mitigation methodologies have already been developed and implemented~\cite{err-mit1,err-mit2,err-mit3}. In this work, we propose a method to address this problem that is complementary to existing approaches and has the potential to significantly improve the accuracy achieved in VQE frameworks~\cite{mcclean2016theory, o2016scalable, kandala2017hardware, romero2018strategies}.

Existing error-mitigation techniques focus on enhancing the computation of expectation values with respect to a parametrized quantum circuit at a single point in parameter space~\cite{temme2017error, err-mit1, err-mit2, wallman2016noise, err-mit3, larose2020mitiq, kim2021scalable, zhang2020error, bonet2018low, mcardle2019error, mcclean2020decoding}. For instance, the zero-noise extrapolation technique measures an observable at a single parameter point for a set of equivalent circuits with different noise strengths, followed by fitting with analytical functions such as polynomials and extrapolating the expectation value to the zero-noise limit~\cite{temme2017error, err-mit1, err-mit2, larose2020mitiq, kim2021scalable}. Probabilistic error cancellation and randomized compiling approaches convert the expectation value with respect to a parametrized circuit at a single point to a sum of estimations with equivalent random circuits, effectively transforming the coherent noise into stochastic error~\cite{temme2017error, wallman2016noise, err-mit3, zhang2020error}.

In contrast to mitigating the error of VQE measurements for each individual variational state, as in the methods mentioned above, we propose a complementary approach that aims to mitigate the error for the entire variational landscape simultaneously by exploiting the underlying correlations between VQE outputs for different variational parameters. Specifically, we employ a custom probabilistic machine learning method based on GPR (rooted in Bayesian statistics)~\cite{RW_gpr}, tailored to incorporate the specific structure of any given parametrized quantum circuit. This method allows us to include within the computation our prior knowledge about general mathematical structure of the variational ansatz.
This approach can be combined with any of the error-mitigation techniques mentioned above, substantially improving the accuracy of VQE calculations.

The manuscript is structured as follows.
In Secs.~\ref{parametrization-subsection}, \ref{GPR-main-section} and \ref{observables-sec} we present our error-mitigation formalism from a general perspective; in Secs.~\ref{practical_example} and \ref{practical_example2} we show benchmark calculations of a Fermionic impurity model and a Heisenberg model using our method within the IBM quantum-computing framework Qiskit~\cite{Qiskit}.

\section{Hamiltonian-independent features of the variational landscape} \label{parametrization-subsection}

Let us consider a generic Hamiltonian $\hat{H}$ and a variational space $\{\ket{\Psi(\btheta)}\}$, where the components of $\btheta=(\theta^{1},..,\theta^{d})$ are real numbers parametrizing the trial quantum states.
Our goal is to determine:
\begin{align}
    \bar{\btheta}&=\text{argmin}_{\btheta}\,\mathcal{E}(\btheta)\,,
    \label{thX}
    \\
    \mathcal{E}(\btheta)&= \Av{\phi(\btheta)}{\hat{H}}
    \label{variationalE}
    \,.
\end{align}

Within VQE frameworks, we prepare the state $\ket{\phi(\btheta)}$ using parametrized quantum gates, and estimate $\mathcal{E}(\btheta)$ from a series of quantum measurements for each $\btheta$. However, outputs from currently available intermediate-scale quantum devices are affected by spurious effects such as decoherence and hardware imperfections, resulting in both random and systematic noise~\cite{cummins2003tackling}.

The key idea at the basis of our method for mitigating the noise in VQE frameworks is to exploit the fact that the variational landscape satisfies exact properties that are known beforehand.
Therefore, corrections to the VQE measurements, which may be affected by spurious noise effects that potentially violate such exact properties, can be rationally enforced for consistency with them. As we are going to show, this approach can effectively mitigate noise by aligning the VQE output with the inherent structure of the variational landscape.
Here we illustrate this point focusing on a generic ansatz represented as follows:
\be
\ket{\phi(\btheta)}=\hat{U}(\btheta)\ket{\phi_0}\,.
\label{ansatz1}
\ee
Here, $\ket{\phi_0}$ is a single-particle state (e.g., the Hartree-Fock solution of $\hat{H}$), and $\hat{U}(\btheta)$ is a unitary transformation given by:
\be
\hat{U}(\btheta)=
\prod_{l=1}^{d} \prod_{m=1}^{M_l} e^{i \hat{G}_{lm} \frac{\theta^{l}}{2}}
\,,
\label{ansatz2}
\ee
where $\theta^{l}$ are variational parameters, the generators $\hat{G}_{lm}$ are Pauli strings expressed in the Hartree-Fock basis, $d$ is the number of variational parameters and $M_l$ is the number of Pauli strings for each variational parameter.
Note that a Pauli string is defined as a generic tensor product of Pauli operators $\hat{P}=\otimes_k\hat{P}^k$, where $\hat{P}^k\in\{\mathds{1}, X^k, Y^k, Z^k\}$,
$k$ is a generic site label, $X^k$, $Y^k$ and $Z^k$ are the corresponding local Pauli operators
and $\mathds{1}$ is the identity.

A key observation is that, as noted in Refs.~\cite{Sequential-VQE,Rotosolve}, since $\hat{G}_{lm}^2 = \hat{I}$ $\forall\,l,m$, where $\hat{I}$ is the identity, we have:
\be
e^{i \hat{G}_{lm} \frac{\theta^{l}}{2} } = \cos\left(\frac{\theta^{l}}{2} \right)\hat{I} + i\sin\left(\frac{\theta^{l}}{2} \right)\hat{G}_{lm}
\,.
\ee
Therefore, the variational energy [Eq.~\eqref{variationalE}] can be expanded in the following form:
\begin{align}
\mathcal{E}(\btheta)&=
\sum_{i_1=0}^{2M_1}...\sum_{i_d=0}^{2M_d}
\xi_{i_1,..,i_d}\,
\prod_{l=1}^d
\cos\left(\frac{\theta^l}{2} \right)^{i_l}
\sin\left(\frac{\theta^l}{2} \right)^{2M_l-i_l}
\nonumber\\
&=\sum_{s=1}^{S} \xi_s T_s(\btheta)\,,
\label{parametric_form}
\end{align}
where we use for convenience a composite index
$s = (i_1,..,i_d)$ to label the coefficients $\xi_s$, 
$T_s(\btheta)$ are the corresponding trigonometric functions, and $S=\prod_{l=1}^d (2M_l + 1)$.
Note that 
the coefficients $\xi_s$ are not fixed by the variational ansatz, but they depend on the specific Hamiltonian operator $\hat{H}$.

It is useful to note that the functions $T_s(\btheta)$  have periodicity $2\pi$, and the condition expressed by Eq.~\eqref{parametric_form} can be conveniently reformulated in terms of the corresponding plane-waves orthonormal basis as follows:
\begin{align}
\mathcal{E}(\btheta) 
&= \sum_{k_1=-M_1}^{M_1}...
\sum_{k_d=-M_d}^{M_d}
\epsilon_{k_1,..,k_d}
\prod_{l=1}^d
\frac{e^{ik_l\theta_l}}{(2\pi)^{1/2}}
\nonumber\\
&= \sum_{\bk}\epsilon_\bk \frac{e^{i\mathbf{k}\mathbf{\btheta}}}{(2\pi)^{d/2}}\,,
\label{parametric_form-k}
\end{align}
where \(\mathbf{k}=(k_1,\ldots,k_d)\) and each \(k_l\) is an integer running from \(-M_l\) to \(M_l\).
Note that this reformulation in terms of the plane-wave basis does not alter the functional space spanned by the \( T_s(\btheta) \) functions, thereby ensuring the exact representation of the variational energy landscape is preserved.
This step shows that Eq.~\eqref{parametric_form} encodes precise information about the smoothness of the energy landscape, guaranteeing that all
Fourier components with $|k_l|>M_l$ are $0$. 
The proposed approach involves utilizing the GPR framework to exploit such prior knowledge about global properties of the variational landscape.
As we are going to show, since Eq.~\eqref{parametric_form-k} correlates the data with each other, it can mitigate  the error arising from individual QPU measurements, and reduce substantially the computational cost of the variational-energy minimization.

\section{GPR-based error mitigation}\label{GPR-main-section}

In this section, we demonstrate how to use the GPR framework in conjunction with Eq.~\eqref{parametric_form} and the variational-energy data $D$ measured on QPUs to compute a "posterior probability distribution" $P[\mathcal{E}|D]$ within the space of energy landscapes.

In the following sections, we show that the resulting posterior probability distribution can be efficiently employed for minimizing the variational energy in VQE frameworks, mitigating errors and significantly reducing the required number of quantum measurements.

\subsection{Prior probability distribution associated with a variational ansatz}\label{parametric_prior-section}

Suppose the Hamiltonian $\hat{H}$ is unspecified, but the variational ansatz [Eq.~\eqref{ansatz1}] has been chosen.
Our aim is to "learn" the energy landscape $\mathcal{E}(\btheta)$ from data $D$ obtained through quantum measurements, and express our prediction in the form of a "posterior probability distribution" $P[\mathcal{E}|D]$.
To construct such a predictive model, we need to incorporate our prior knowledge about the function to be learned (i.e., the information available before any measurement) into the calculation.

In the GPR framework, this prior knowledge is encapsulated in a "prior probability distribution" $P[\mathcal{E}]$, which is conveniently assumed to be Gaussian.
A significant benefit of supposing that $P[\mathcal{E}]$ is Gaussian is that all Gaussian distributions are entirely characterized by their corresponding "propagator" (or Kernel function), which is defined as follows:
\begin{align}
K(\btheta_1,\btheta_2)=\int \D[\mathcal{E}] P[\mathcal{E}]\, \mathcal{E}(\btheta_1)\,\mathcal{E}(\btheta_2)
\,,
\label{kernel-definition}
\end{align}
where $\D[\mathcal{E}]$ is the standard functional-integral measure (see the Supplemental material~\cite{supplemental_material} for a more detailed introduction to GPR).
A key feature of the proposed approach is that, unlike in standard GPR where the kernel function is typically chosen empirically from a list (including, e.g., the widely used "square exponential" kernel), here we \emph{calculate} it from Eq.~\eqref{parametric_form}, 
as follows:
\begin{align}
P[\mathcal{E}] 
 \!\propto\!\! \int\! \prod_r d\xi_r & 
\exp\!\left\{\!-\frac{1}{2\eta^2}\!\!
\int\!\! {\bdtheta}\, \left[\mathcal{E}({\btheta})\!-\!\sum_s\xi_s T_s({\btheta})\right]^2
\!\right\}
\nonumber
\\
\times & \exp\!\left\{\!-\frac{t}{2}\!\!
\int\!\! {\bdtheta}\, \mathcal{E}^2({\btheta})
\!\right\}
\label{parametric_prior}
\,,
\end{align}
where:
\be
{\bdtheta}=\prod_{l=1}^d {d\theta}^l\,,
\ee
the integrals over $d\theta^l$ are all taken between $-\pi$ and $\pi$,
and $\eta$ is considered in the limit $\eta \rightarrow 0$.
The parameter $t$ encodes our prior information concerning the range of the variational energy, as the second factor in Eq.~\eqref{parametric_prior} vanishes exponentially for $|\mathcal{E}(\btheta)|\gg t^{-1/2}$.

The meaning of Eq.~\eqref{parametric_prior} is that,
within our scenario, we know a priori that $\mathcal{E}(\btheta)$ can be expressed as in Eq.~\eqref{parametric_form}, where $T_s(\btheta)$ are known analytical functions. However, we lack prior information about the coefficients $\xi_s$, except for the condition that the resulting energy landscape must be bounded for all physical Hamiltonians.
As demonstrated in the supplemental material~\cite{supplemental_material}, the functional integral in Eq.~\eqref{parametric_prior} can be calculated explicitly, leading to:
\begin{align}
    K(\btheta_1,\btheta_2)&=
    \frac{t^{-1}}{(2\pi)^d}
    \sum_\bk e^{i \bk(\btheta_1-\btheta_2)}
    \label{kernel-explicit}
    \\
    \nonumber
    &=\frac{t^{-1}}{(2\pi)^d}\sum_{k_1=-M_1}^{M_1}...
    \sum_{k_d=-M_d}^{M_d} e^{i \bk(\btheta_1-\btheta_2)}
    \,.
\end{align}


It should be noted that the framework above has the flexibility of encoding in the prior also additional information that may be available to us for specific VQE.
As an example, in Sec.~\ref{practical_example2} we consider the example of a Heisenberg model such that, because of the specific generators $\hat{G}_{lm}$ used, the unitary transformation $\hat{U}(\btheta)$ has periodicity $\pi$ instead of $2\pi$.
Since, under such hypothesis, only the coefficients $\epsilon_\bk$
of Eq.~\eqref{parametric_form-k} with even $k_l$ are non-zero, we have that:
\begin{align}
    K(\btheta_1,\btheta_2)&=
    \frac{t^{-1}}{(2\pi)^d}
    \sum_\bk e^{i \bk(\btheta_1-\btheta_2)}
    \label{kernel-explicit-2}
    \\
    \nonumber
    &=\frac{t^{-1}}{(2\pi)^d}\sum_{k_1=-M_1/2}^{M_1/2}...
    \sum_{k_d=-M_d/2}^{M_d/2} e^{i 2\bk(\btheta_1-\btheta_2)}
    \,.
\end{align}

\subsection{QPUs data}\label{QPUsdata}

Let us consider any fixed Hamiltonian $\hat{H}$ whose energy landscape is a function $\mathcal{E}(\btheta)$ sampled from our prior probability distribution $P[\mathcal{E}]$.

For each variational parameter $\btheta_\alpha$ we consider the random variable:
\begin{align}
    \mathcal{E}_\alpha=\frac{1}{N}\sum_{r=1}^{N}\mathcal{E}_{\alpha r}
    \,,
\end{align}
describing the average of a series of $N$ quantum measurements (shots), given the variational state $\ket{\phi(\btheta_\alpha)}$ prepared on the quantum circuit.
We assume that, for large $N$, the conditional probability of obtaining $\mathcal{E}_\alpha$ can be represented as follows:
\begin{align}
P(\btheta_\alpha|\mathcal{E}_\alpha)\propto
\exp\left\{-\frac{1}{2\sigma^2_\alpha}
(\mathcal{E}(\btheta_\alpha)-\mathcal{E}_\alpha)^2\right\}
\,,
\label{PDf-step}
\end{align}
i.e., that $\mathcal{E}_\alpha$ is approximately gaussian distributed around the real underlying variational energy:
\begin{align}
\mathcal{E}(\btheta_\alpha)=\Av{\phi(\btheta_\alpha)}{\hat{H}}
\label{Eformal}
\,,
\end{align}
with a variance $\mathcal{\sigma}_\alpha$.

We note that Eq.~\eqref{PDf-step} is rigorously applicable primarily to ideal (fault-free) quantum machines. In fact, in such scenarios the outcome of quantum measurements is inherently probabilistic, and the variance \( \mathcal{\sigma}^2_\alpha \) can be precisely defined as \( \Av{\phi(\btheta_\alpha)}{\hat{H}^2-\mathcal{E}^2(\btheta_\alpha)} / {N} \).
In contrast, on real quantum devices \( \sigma_\alpha \) must be regarded as an aggregate measure of the "error bar" for quantum measurements. This measure encompasses not only the intrinsic uncertainties of quantum projective measurements but also includes spurious effects like decoherence and hardware imperfections. 
Consequently, \( \sigma_\alpha \) is effectively a characteristic of the specific quantum device and is thus treated as a "hyperparameter" within our GPR framework. This hyperparameter plays a crucial role in guiding the GPR model in determining how closely the posterior probability distribution for the variational landscape should align with the energy values measured by the quantum device.

It's important to highlight that this setting is typical in the practical application of GPR, where scenarios featuring an error bar that is intrinsically probabilistic and Gaussian are rarely, if ever, encountered. In real-world applications, GPR is commonly utilized in contexts where the error bar reflects a combination of inherent uncertainties and additional, device-specific noise factors.

In summary, we assume that the probability of obtaining a data set:
\be
D=\{(\btheta_\alpha,\mathcal{E}_\alpha,\sigma_\alpha)\,|\,\alpha=1,..,n\}\,, 
\label{data}
\ee
given an underlying Hamiltonian $\hat{H}$ and a variational
state $\ket{\phi(\btheta_\alpha)}$, can be estimated by the following equation:
\begin{align}
P[D|\mathcal{E}]&\propto
\prod_{\alpha=1}^n
P(\btheta_\alpha|\mathcal{E}_\alpha)
\nonumber
\\
&\propto
\exp\left\{-\sum_{\alpha=1}^n 
\frac{1}{2\sigma^2_\alpha}
(\mathcal{E}(\btheta_\alpha)-\mathcal{E}_\alpha)^2
\right\}
\label{PDf}
\,,
\end{align}
where $\sigma_\alpha$ is a hyperparameter (that depends on the
"quality" of the quantum device) for all $\btheta_\alpha$
such that the energy is evaluated from projective measurements.

\subsection{The posterior probability distribution}

Applying Bayes' rule with the given data $D$ (refer to Eq.~\eqref{data}) and a prior probability $P[\mathcal{E}]$, we obtain the following "posterior" conditional probability for the function $\mathcal{E}$:
\be
P[\mathcal{E}|D]\propto P[\mathcal{E}] 
\exp\left\{-\sum_{\alpha=1}^n 
\frac{1}{2\sigma^2_\alpha}
(\mathcal{E}(\btheta_\alpha)-\mathcal{E}_\alpha)^2
\right\}
\,,
\label{posterior}
\ee
which represents the probability distribution for the function $\mathcal{E}$, given a data set $D$, see Eq.~\eqref{data}.

The interpretation of Eq.~\eqref{posterior} is that the probability of a specific energy landscape is influenced by the prior distribution and "anchored" to the data, i.e., it is exponentially suppressed for configurations deviating from the measured data $\mathcal{E}_\alpha$ by more than $\sigma_\alpha$.

Since $P[\mathcal{E}|D]$ is Gaussian, we can calculate the following quantities exactly:
\begin{align}
\langle \mathcal{E}(\btheta) \rangle &= \int \D[\mathcal{E}] P[\mathcal{E}|D]\, \mathcal{E}(\btheta)
\label{av}
\\
\Sigma^2(\btheta)  &= \int \D[\mathcal{E}] P[\mathcal{E}|D]\, \big(\mathcal{E}^2(\btheta)-\langle \mathcal{E}(\btheta) \rangle^2\big)
\label{sq}
\,,
\end{align}
where Eq.~\eqref{av} represents our prediction for $\mathcal{E}(\btheta)$ at any test point $\btheta$ and Eq.~\eqref{sq} represents the uncertainty of our prediction.

Let us express explicitly $\langle \mathcal{E}(\btheta) \rangle$ and $\Sigma(\btheta)$
as a function of the data $D$ (refer to Eq.~\eqref{data}).
We define the matrix:
\begin{align}
\bar{K}_{\alpha\beta}=K(\btheta_\alpha,\btheta_\beta)+
\sigma^2_\alpha\delta_{\alpha\beta}\quad
\forall\,\alpha,\beta\in 1,..,n
\,,
\end{align}
where $\delta_{\alpha\beta}$ is the Kroneker delta, and $K$ is the kernel function defined in Eq.~\eqref{kernel-definition} (which is given by Eq.~\eqref{kernel-explicit} for our prior probability distribution).
Following the procedure detained in the supplemental material~\cite{supplemental_material} for completeness, we have:
\begin{align}
\langle \mathcal{E}(\btheta) \rangle &= 
\sum_{\alpha,\beta=1}^n
K(\btheta,\btheta_\alpha)
[\bar{K}^{-1}]_{\alpha\beta}\mathcal{E}_\beta
\label{av-explicit}
\\
\Sigma^2(\btheta)  &= 
K(\btheta,\btheta)-
\sum_{\alpha,\beta=1}^n
K(\btheta,\btheta_\alpha)
[\bar{K}^{-1}]_{\alpha\beta}
K(\btheta_\beta,\btheta)
\,.
\label{sq-explicit}
\end{align}

Note that evaluating Eqs.~\eqref{av-explicit} and \eqref{sq-explicit} requires computing the inverse of $\bar{K}$, whose size equals the number $n$ of training data points.
Therefore, for our approach to be practically applicable, it is important that $n$ does not become prohibitively large.
On the other hand, 
minimizing the variational energy does \emph{not} necessitate learning the entire variational landscape (which is encoded in $S$ parameters, see Eq.~\eqref{parametric_form}).
In fact, as we are going to show in our benchmark calculations, the number $n$ of training data point necessary for this task is generally smaller than $S$.

In the next section we describe a general method for calculating the VQE energy minima from QPU measurements, by combining the GPR approach framework discussed above with standard minimization algorithms.

\begin{figure}
    \centering
    \includegraphics[width=0.45\textwidth]{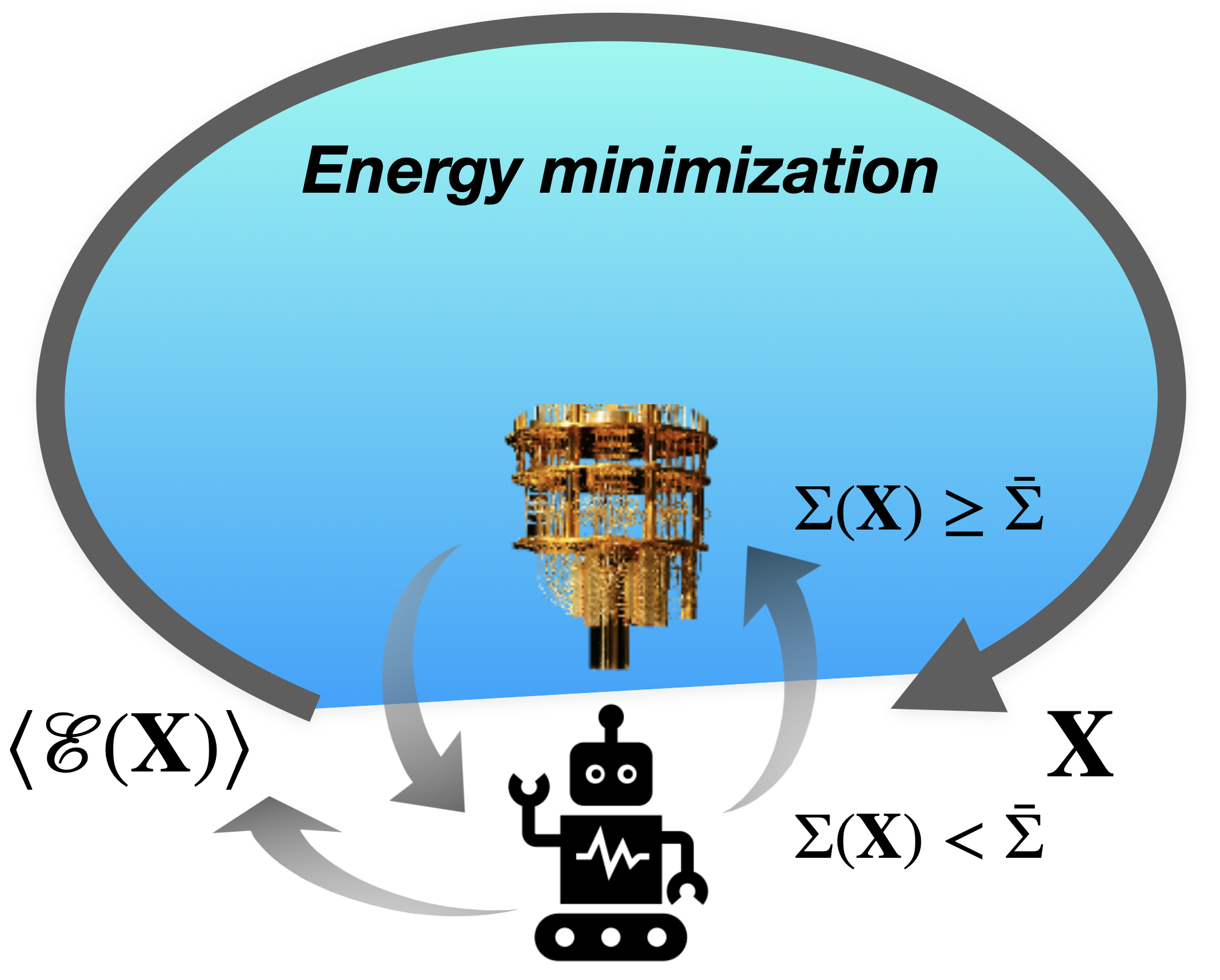}
    \caption{Schematic illustration of the proposed energy minimization algorithm based on GPR. The method incorporates a GPR machine with a customized prior based on the specific VQE ansatz. The main steps of the algorithm are shown, including evaluation of uncertainty, utilization of GPR estimates when the uncertainty is below a threshold, and updating the GPR database when the uncertainty is above the threshold.}
   \label{Figure1}
\end{figure}

\section{Energy minimization with active learning framework}\label{observables-sec}

The standard approach for computing the minimum of the variational energy $\mathcal{E}(\btheta)$ involves evaluating the energy for a series of different variational parameters by averaging over a series of quantum measurements at each iteration. This method does not take advantage of any prior knowledge about the variational ansatz.

In contrast, we propose a modified procedure, as illustrated in Fig.~\ref{Figure1}, which incorporates the GPR method and a customized prior based on the specific variational ansatz implemented on the parametrized quantum circuit. This new algorithm can be summarized as follows.
After initializing the posterior probability distribution starting from an empty dataset $D$, we perform the following steps:
\begin{itemize}

\item Whenever the minimization algorithm requests to evaluate the variational energy for a given set of variational parameters $\btheta$, compute the uncertainty $\Sigma(\btheta)$ using GPR, based only on the information already available in $D$.

\item If $\Sigma(\btheta)\leq \bar{\Sigma}$, where $\bar{\Sigma}$ is a predefined accuracy threshold, proceed using the GPR estimate for the variational energy $\langle \mathcal{E}(\btheta)\rangle$, without performing any additional quantum measurement.

\item If $\Sigma(\btheta)>\bar{\Sigma}$, estimate the variational energy using quantum measurements, and load this information into the GPR database $D$. Then, provide the obtained GPR estimate $\langle \mathcal{E}(\btheta)\rangle$ to the minimization algorithm.
\end{itemize}

Note that our procedure, as outlined above, departs from the standard Bayesian optimization, which typically employs GPR uncertainty quantification to guide exploration in the variational space (a strategy generally feasible only for low-dimensional problems). Our approach, in contrast, relies on an independent energy minimization procedure, and uses GPR only to bypass QPU evaluations when possible. In this general sense, our algorithm's structure bears more resemblance to the adaptive scheme employed in building potential energy surfaces (PESs) using GPR, as seen in Ref.~\cite{schmitzGaussianProcessRegression2020}.

A crucial element of our framework is the custom kernel function, crafted to match the structure of the VQE circuit, thus improving GPR predictions by incorporating the specific form of the variational ansatz in the prior.
Additionally, it is essential to point out that our approach does not aim to learn the full VQE energy landscape. Instead, it targets the trajectory of the energy-minimization process, which is essentially a one-dimensional path in the parameter space, from the start to the energy minimum. 
Therefore, our method effectively avoids the computational issues often associated with high-dimensional spaces, like the curse of dimensionality. 
In particular, as the minimization process advances, the explored points become increasingly close to each other, making our prior knowledge about the landscape's smoothness increasingly relevant, regardless of the number of variational parameters in the VQE ansatz.

In the next section, we illustrate in detail the methodology proposed above for calculations of a series of 2-qubit impurity-model Hamiltonians and present benchmark calculations performed on real quantum devices.

\section{Benchmark calculations of a 2-qubit impurity model}\label{practical_example}

We consider a Fermionic Hamiltonian represented as follows:
\begin{align}
\hat{H} &= \frac{U}{2}(\hat{n}_c-1)^2
+\mathcal{D}\!\sum_{\sigma=\uparrow,\downarrow} \!
\left(\cc_{\sigma}\da_{\sigma}+\dc_{\sigma}\ca_{\sigma}\right)
\nonumber
\\
&+\lambda^c\!\!\sum_{\sigma=\uparrow,\downarrow} \da_{\sigma}\dc_{\sigma}
\,,
\label{hemb}
\end{align}
where: $\cc_\sigma$ and $\ca_\sigma$ are the creation and annihilation operators of the so-called "impurity degrees of freedom", $\dc_\sigma$ and $\da_\sigma$ are the creation and annihilation operators of the so-called "bath degrees of freedom", $\sigma\in\{\uparrow,\downarrow\}$ is the spin index, $\hat{n}_c=\sum_\sigma\cc_{\sigma}\ca_{\sigma}$ is the impurity number operator, $U$ is the Hubbard-repulsion parameter for the impurity degrees of freedom, while $\lambda^c$ and $\mathcal{D}$ are the coupling constants characterizing the bath of the impurity model and its coupling to the impurity, respectively.
From now on we set $\D=-\vert \D\vert$, with $|\D|$ serving as the unit of energy.

We emphasize that Eq.~\eqref{hemb} represents the simplest possible "impurity model" or "embedding Hamiltonian" (EH), a fundamental building block of QE methods. Given that solving the EH in QE calculations is one of the most promising potential applications of VQE, our choice of this model for benchmark calculations aims to showcase the potential of our method in addressing more complex quantum systems typically encountered in practical QE calculations.

Specifically, we illustrate and benchmark our method for calculating
the ground state $\ket{\Phi(\mathcal{D},\lambda^c,U)}$ of $\hat{H}$, within
the subspace generated by states with $2$ electrons (i.e., half of the
maximum possible occupation).
The QPU data are generated using the IBM open-source framework for quantum computing Qiskit~\cite{Qiskit}, which provides methods for manipulating quantum programs on real quantum computers, as well as on classical QPU simulators.

\subsection{Qubit representation of the EH}

Following the approach in Ref.~\cite{YY-qc}, we transform the EH (see Eq.~\eqref{hemb}) using the so-called "parity mapping"~\cite{Bravyi-Kitaev}, as implemented in Qiskit~\cite{Qiskit}, leading to the following 2-qubit representation:
\begin{align}
\hat{H}&=\zeta_0\mathds{1}+
\zeta_1(\sigma^z_1-\sigma^z_2)+\zeta_2(\sigma^x_1+\sigma^x_2)+\zeta_3 \sigma^z_1\sigma^z_2
\nonumber
\\
&+\zeta_4(\sigma^x_1\sigma^z_2-\sigma^z_1\sigma^x_2)+\zeta_5 \sigma^x_1\sigma^x_2\,,
\label{hemb-spins}
\end{align}
where the symbols $\sigma^x_k$, $\sigma^y_k$, and $\sigma^z_k$ (with $k = 1, 2$) represent the Pauli matrices acting on the $k$-th qubit, and
the coefficients $\zeta_j$ are determined 
from the EH parameters $U$, $\D$ and $\lambda^c$ through the parity mapping.

\subsection{The VQE ansatz}\label{trigonometric-expansion}

To calculate the spin-singlet ground state of the EH we use an ansatz inspired by unitary coupled cluster ansatz with single and double excitations (UCCSD) ~\cite{o2016scalable,UCCSD2,UCCSD3, UCCSDrev}.We here express the wave function directly in qubit basis similar to previous works\cite{o2016scalable, YY-qc} and expressed as products of unitary single and double rotations:
\begin{align}
    \ket{\phi(\theta)}=
    e^{\frac{i}{2}\theta^1 \sigma^y_2}e^{-\frac{i}{2}\theta^1\sigma^y_1}e^{-\frac{i}{2}\theta^2 \sigma^y_1\sigma^x_2}
    e^{\frac{i}{2}\theta^2 \sigma^x_1\sigma^y_2}\,
    \ket{\phi_0}
    \,,
    \label{special_case}
\end{align}
Here $\theta=(\theta^1,\theta^2)$, with each angle in the range of $[-\pi,\pi]$ and $\ket{\phi_0}$ is the spin-restricted Hartree-Fock ground-state solution of $\hat{H}$. The angles for the single-qubit rotations are set to be equal because of spin rotational symmetry.

Note that Eq.~\eqref{special_case} is a special case of Eqs.~\eqref{ansatz1} and \eqref{ansatz2},
with $d=2$ and $M_1=M_2=2$.
Therefore, from the framework described in Sec.~\ref{parametrization-subsection} it follows that the variational energy $\mathcal{E}(\theta)$ can be expressed as a linear combination of 25 trigonometric functions of the angles:
\begin{align}
    \mathcal{E}(\btheta) &= 
    \sum_{k_1,k_2=-2}^2
    \epsilon_{k_1,k_2} \frac{e^{i\bk \cdot \btheta}}{2\pi}
    \,,
\end{align}
and the GPR framework can be applied with the Kernel function:
\begin{align}
    K(\btheta_1,\btheta_2)&=
    \frac{t^{-1}}{(2\pi)^2}
    \sum_{k_1,k_2=-2}^2
    e^{i \bk \cdot (\btheta_1-\btheta_2)}
    \label{kernel-explicit3}
    \,.
\end{align}
The variational energy estimation $\langle \mathcal{E}(\btheta) \rangle$ and uncertainty quantification $\Sigma(\btheta) $ are given by Eqs.~\eqref{av-explicit} and \eqref{sq-explicit}, respectively.

\begin{figure*}
    \centering
    \includegraphics[width=15.8cm]{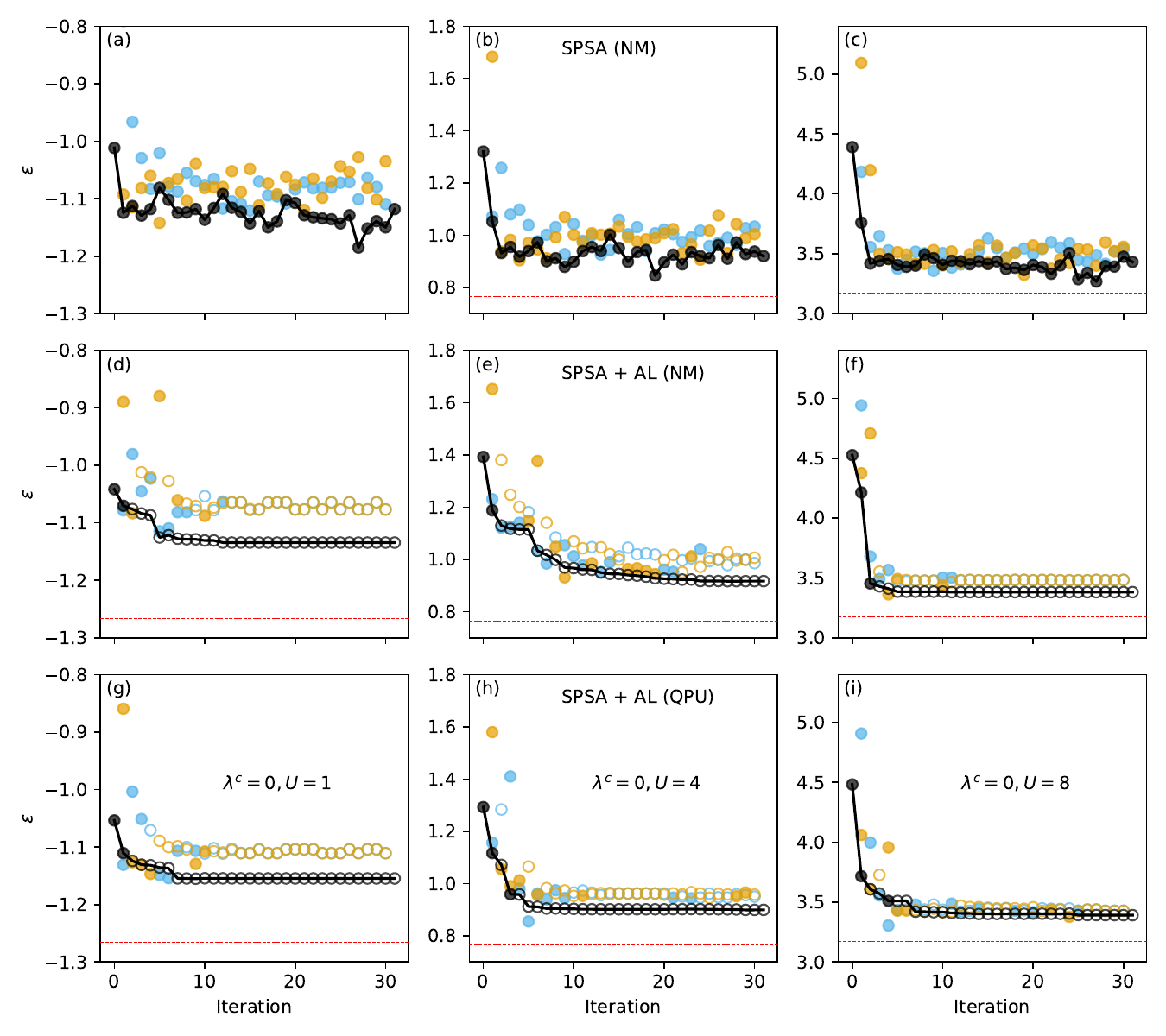}
    \caption{Comparison of energy convergence for the impurity model as a function of iteration steps in the energy minimization procedure, using SPSA and SPSA+AL optimization methods on simulator with noise model (NM) derived from \textsc{ibmq\_mumbai} and \textsc{ibmq\_mumbai} QPU.
    Upper panels a, b, and c show the results for SPSA (NM), middle panels d, e, and f for SPSA+AL(NM), and lower panels g, h, and i for SPSA+AL(QPU).
    For the impurity model, we fix $\lambda^c=0$ and vary Hubbard interaction strength $U=1$ (left panels), $U=4$ (middle panels) and $U=8$ (right panels), representing weakly interacting to strongly correlated regimes. The red dashed horizontal lines indicate the exact state vector results. 
    Full symbols indicate points obtained from direct energy measurement and added to GPR dataset $D$, while empty symbols represent points from GPR prediction without requiring additional measurement. Blue and orange circles indicate points used for gradient calculations, while black circle represents energy reached at each iteration. 
    With 30 iterations, the number of times to measure the energy $\mathcal{E}(\btheta)$ is $92$ for SPSA, and falls in the range of $[15, 27]$ for SPSA+AL.
    }
   \label{Figure2}
\end{figure*}

\begin{table}[t]
  \centering
    \begin{tabular}{|c|c|>{\raggedleft\arraybackslash}p{0.5in}|>{\raggedleft\arraybackslash}p{0.5in}|>{\raggedleft\arraybackslash}p{0.5in}|>{\raggedleft\arraybackslash}p{0.5in}|>{\raggedleft\arraybackslash}p{0.5in}|}
  \hline
  \multirow{2}{*}{$\lambda^c$} & \multirow{2}{*}{$U$} & \multicolumn{2}{c|}{\textbf{SPSA(NM)}} & \multicolumn{2}{c|}{\textbf{SPSA+AL(NM)}} & \multirow{2}{*}{\textbf{EXACT}}\\\cline{3-6}
  & & \textbf{AVG} & \textbf{STD} & \textbf{AVG} & \textbf{STD}  & \\\cline{1-7}
  \multirow{3}*{0} & 1 & -1.121 & 0.016 & -1.134 & 0.009 & -1.266 \\\cline{2-7}
  & 4 & 0.921 & 0.046 & 0.917 & 0.016 & 0.764 \\\cline{2-7}
  & 8 & 3.431 & 0.049 & 3.386 & 0.025 & 3.172 \\\hline
  \multirow{3}*{1} & 1 & -2.295 & 0.019 & -2.311 & 0.015 & -2.454 \\\cline{2-7}
  & 4 & -0.164 & 0.041 & -0.174 & 0.021 & -0.323 \\\cline{2-7}
  & 8 & 2.376 & 0.055 & 2.355 & 0.034 & 2.140 \\\hline
  \multirow{3}*{2} & 1 & -3.782 & 0.029 & -3.797 & 0.014 & -3.968 \\\cline{2-7}
  & 4 & -1.420 & 0.050 & -1.440 & 0.023 & -1.604 \\\cline{2-7}
  & 8 & 1.314 & 0.078 & 1.252 & 0.036 & 1.038 \\\hline
  \end{tabular}
  \vspace{0.1cm}
  \caption{Average and standard deviation of ground-state energies obtained using SPSA(NM) and SPSA+AL(NM) optimization methods of the impurity model over a sample of 20 runs each with different bath energy levels ($\lambda^c = 0, 1, 2$) and Hubbard interaction strengths ($U=1, 4, 8$). Each row corresponds to a specific combination of U and $\lambda^c$. Columns present the average and standard deviation for SPSA(NM), and SPSA+AL(NM) results, with the last column displaying the exact state vector energies for comparison.}
  \label{tab:1}
\end{table}

\subsection{Numerical tests on QPU simulators and real quantum devices}\label{real-machine-benchmarks}

In this subsection, we evaluate the performance of our parametric GPR method for the impurity model described by Eqs.~\eqref{hemb} and \eqref{hemb-spins}. We consider bath parameters $\lambda^c=0,1,2$ and Hubbard interaction strengths $U=1,4,8$, covering a wide range of interaction scenarios, from weakly interacting to strongly correlated regimes.

To demonstrate the effectiveness of our method, we used the Simultaneous Perturbation Stochastic Approximation (SPSA)~\cite{spall1998overview}, a standard optimization approach used for problems involving noisy function evaluations, comparing the results obtained from the bare method with the results augmented with an active learning strategy, here referred to as SPSA+AL. 
In our SPSA+AL calculations, we set the following hyperparameters for the active learning framework: $\sigma_\alpha = \bar{\Sigma} = 0.005$ and $t=1/(2\pi)^2$ such that $K(\btheta, \btheta) = K_c = 25$. We note that $K_c$ is larger than the numerical estimation of it by sampling a set of $N=100$ random angles $\{\btheta_\alpha\}$, $\frac{1}{N}\sum_{\alpha=1}^{N}[\mathcal{E}(\btheta_\alpha)]^2$, for the model at all the parameter points, which falls in the range of $(0.5, 15)$. 
%
We chose $\bar{\Sigma} = \sigma_\alpha$ to align the accuracy threshold for the GPR uncertainty, used in the active learning strategy, with the inherent limitations in the accuracy of the training data points. 
Although we focus on these specific hyperparameter values for the paper, we have also tested other values, such as $\sigma_\alpha = \bar{\Sigma} = 0.05$ or $\sigma_\alpha = \bar{\Sigma}/2 = 0.005$, and found our results to be qualitatively consistent, indicating the robustness of our method.
All points have been generated from either QPU simulators or real quantum devices and loaded within the GPR framework, following the active learning procedure described in Sec.~\ref{observables-sec}.

Figure 2 illustrates the energy evolution as a function of the number of iteration steps throughout the energy minimization procedure, employing bare SPSA (panels a, b, and c) and SPSA+AL (panels d, e and f) for $\lambda^c=0$ and $U=1, 4, 8$. 
Panels g, h, and i show the energy convergence of SPSA+AL for real quantum devices (QPU) across the range of interaction strengths, $U=1, 4, 8$.

Our results show that, SPSA+AL exhibits rapid convergence compared with bare SPSA across all parameter regimes, necessitating a significantly smaller number of QPU energy evaluations. 
This holds true for both the data produced by the simulator and the data obtained from real quantum devices, demonstrating the ability of our method to adapt to the distinct noise characteristics of real hardware.

To systematically compare the performance of bare SPSA and SPSA+AL, we compiled the results in Table 1, which presents the average and standard deviation of ground-state energies obtained for 30 energy minimization steps of the impurity model with different Hubbard interaction strengths ($U=1, 4, 8$) and bath energy levels ($\lambda^c = 0, 1, 2$). For both SPSA and SPSA+AL, the energy minimum evaluations were computed after 30 iterations, as the energy oscillated without further improvement beyond that point, as illustrated in Fig.~2. These results expand the analysis above in two significant ways. Firstly, by including 20 repetitions of the simulation, we demonstrate the robust performance of our SPSA+AL method. Secondly, by extending the analysis away from half-filling, we showcase the effectiveness of our approach for non-zero values of $\lambda^c$.

It should be noted that the data presented in Table 1 are derived exclusively from simulations, due to the limited availability of real quantum devices. However, the successful application of our method to real device data in the analysis of Figure 2 suggests that the conclusions drawn from the simulated data remain valid and applicable to real quantum hardware.

The analysis presented in Table 1 highlights the superior performance of SPSA+AL compared to bare SPSA.
For each iteration, SPSA queries two times for the value of $\mathcal{E}(\btheta)$ to estimate the gradient, and one time for the energy at the current step, which allows to revert the change of variational angles if the updated energy shoots over a preset threshold ($0.5$ for current calculations). Therefore, the total 30 SPSA iterations correspond to $30\times3+2=92$ times (including the initial and final energy measurement) of direct energy measurement for each run. In contrast, SPSA+AL requires only 15 to 27 times for energy measurement per run. Furthermore, SPSA+AL consistently delivers slightly more accurate energies for all values of $U$ and $\lambda^c$ considered.
This demonstrates the effectiveness of our approach in enhancing the optimization process, reducing the number of QPU evaluations, and showcasing its robustness and applicability across different regimes.
It is worth noting that the standard deviations observed for SPSA+AL method are systematically lower than the other methods, indicating reduced fluctuations compared to SPSA. This reduction in fluctuations could potentially improve the stability of QE methods using VQE as impurity solvers, as such methods require iterative impurity-model solutions with varying parameters, and output fluctuations can hamper stability and convergence.

\begin{figure}
    \centering
    \includegraphics[width=0.45\textwidth]{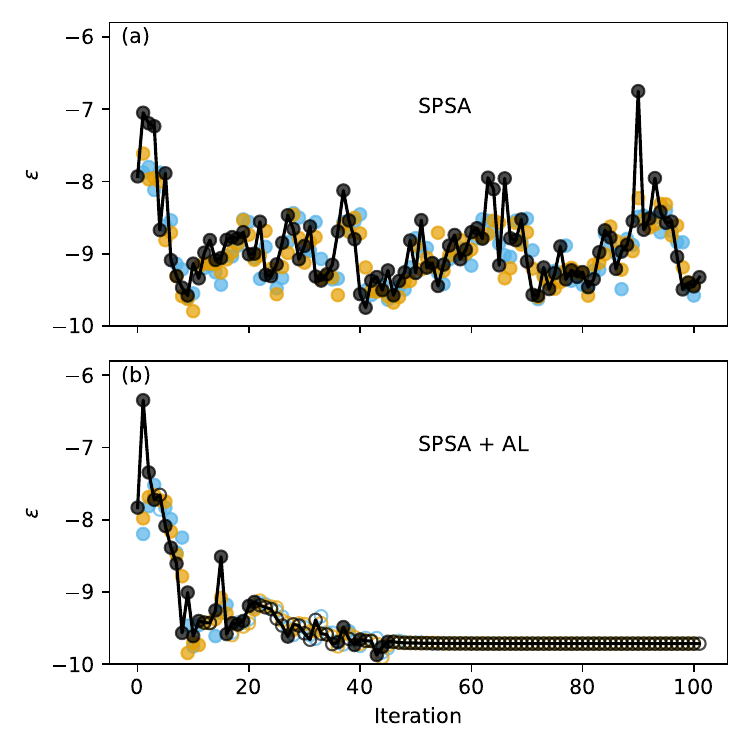}
    \caption{Comparison of energy convergence for the Heisenberg model as a function of iteration steps in the energy minimization procedure, using SPSA and SPSA+AL optimization methods on simulator with noise model derived from \textsc{ibmq\_mumbai} QPU.
    Panel (a) shows the results with SPSA, while panel (b) shows that with SPSA+AL.
    We use the same color encoding as Fig.~\ref{Figure2} for the data points: full symbols from direct energy measurement, while empty symbols from GPR prediction. Blue and orange circles indicate points used for gradient calculations, while black circle represents energy at each iteration. With 100 iterations, the number of times to measure the energy $\mathcal{E}(\btheta)$ is $302$ for SPSA, and $61$ for SPSA+AL. The simulation is performed for 8-site Heisenberg chain with antiferromagnetic coupling.}
   \label{Figure3}
\end{figure}

\section{Benchmark calculations of a 8-qubit Heisenberg model}\label{practical_example2}
To further demonstrate the effectiveness of SPSA+AL approach, we apply it to variational ground state preparation of an $L$-site Heisenberg chain with Hamiltonian:
\be
\h = J\sum_{i=1}^{L-1}\boldsymbol{\sigma}_i\cdot \boldsymbol{\sigma}_{i+1}. 
\ee
Here $\boldsymbol{\sigma}_i$ is the vector of Pauli operators at site $i$, and we set $J=1$ for antiferromagnetic coupling.

\subsection{The VQE ansatz}\label{trigonometric-expansion2}
The following Hamiltonian variational ansatz is adopted for ground state preparation~\cite{ho2019efficient}:
\be
\ket{\phi(\btheta)} = e^{\frac{i}{2}\theta^1 H_1} e^{\frac{i}{2}\theta^2 H_2} \ket{\phi_0} \label{eq: hva},
\ee
with $H_1 = \sum_{i=1}^{L/2} \boldsymbol{\sigma}_{2i-1}\cdot \boldsymbol{\sigma}_{2i}$ and $H_2 = \sum_{i=1}^{L/2-1} \boldsymbol{\sigma}_{2i}\cdot \boldsymbol{\sigma}_{2i+1}$. The reference state $\ket{\phi_0}=\otimes_{i=1}^{L/2} \frac{1}{\sqrt{2}}(\ket{\uparrow\downarrow}-\ket{\downarrow\uparrow})_{2i-1, 2i}$ is a product state of Bell pairs (spin singlets), which is also the ground state of $H_1$. Each angle of $\btheta$ can be restricted to range of $[-\pi/2, \pi/2]$. 

For the following numerical simulations, we set $L=8$, corresponding to 8 qubits. The ansatz of Eq.~\eqref{eq: hva} corresponds to $d=2$, $M_1=12$, and $M_2=9$ for generic expression of Eqs.~\eqref{ansatz1} and \eqref{ansatz2}.
The variational energy $\mathcal{E}(\btheta)$ can therefore be expressed as a linear combination of $25\times 19$ trigonometric functions of the angles:
\begin{align}
    \mathcal{E}(\btheta) &= 
    \sum_{k_1=-12}^{12}\sum_{k_2=-9}^{9}
    \epsilon_{k_1,k_2} \frac{e^{i\bk \cdot \btheta}}{2\pi}
    \,,
\end{align}
and the GPR framework can be applied with the Kernel function:
\begin{align}
    K(\btheta_1,\btheta_2)&=
    \frac{t^{-1}}{(2\pi)^2}
    \sum_{h_1=-6}^{6}\sum_{h_2=-4}^{4}
    e^{i 2\mathbf{h} \cdot (\btheta_1-\btheta_2)}
    \label{kernel-explicit2}
    \,,
\end{align}
where we take into account that the period of $\mathcal{E}(\btheta)$ with respect to $\btheta$ is $\pi$ rather than $2\pi$. This originates from the special form of the generators of $H_1$ and $H_2$, where we have $\prod_{\mu\in \setof{x, y, z}}\sigma^{\mu}_{i}\sigma^{\mu}_{i+1} \propto \mathds{1}$, with $\sigma^{\mu}$ labeling the three components of Pauli matrices and $\mathds{1}$ for the identity.

\subsection{Numerical tests on QPU simulators }\label{real-machine-benchmarks2}

In Fig.~\ref{Figure3}, we present an energy versus iterations plot for the 8-site Heisenberg model using the SPSA and SPSA+AL optimization methods, illustrating a representative example of how a standard run proceeds.
In line with our findings from the impurity model calculations, the SPSA+AL method exhibits 
significantly enhanced convergence efficiency, requiring on average only about 20\% of the energy 
measurements compared to the bare SPSA method. To quantify this, we conducted an ensemble analysis 
involving 20 separate runs for each of the SPSA and SPSA+AL simulations. The results show that the 
averaged final energy for SPSA is $-9.321\pm 0.311$, while for SPSA+AL, it is notably lower at 
$-9.701\pm 0.151$, indicating a more precise convergence with a reduced variance. It's important to  note, however, that the exact ground state energy for this model is $-13.299$. This significant 
discrepancy points to substantial biases in both simulation methods, consistent with the trends 
observed in the impurity model calculations.

While the primary objective of the AL method is not to directly address systematic errors, it plays a 
crucial role in ensuring consistency with the pre-established mathematical structure of the variational landscape. 
To evaluate the efficacy of this indirect form of error mitigation, we calculate the fidelity of the converged solution, defined as \(f=\left|\langle \phi(\btheta)|\,\psi_\text{G}\rangle \right|^2\) with 
respect to the exact ground state \(\ket{\psi_\text{G}}\), utilizing the optimized angles \(\btheta\). 
From the same sample set, we observe a fidelity of \(f=0.888\pm 0.051\) for SPSA and a notably higher 
\(f=0.945 \pm 0.016\) for SPSA+AL, demonstrating enhanced fidelity and reduced variance. Remarkably, the 
mean fidelity of \(f=0.945\) closely approaches the maximum achievable fidelity of \(f=0.960\) as per the 
ansatz in Eq.~\eqref{eq: hva}. This suggests that, while our AL method primarily enforces consistency in 
the variational landscape, conventional error mitigation techniques might still be beneficially applied 
post-SPSA+AL optimization to further diminish measurement biases, aligning with observations in existing 
literature~\cite{mukherjee2023comparative}.

It is noteworthy that in the case of our impurity model calculations, which involves a simpler two-qubit system, the fidelities of the final variational states consistently exceeds $99.7\%$. This observation suggests the increased value of the exact information incorporated into the variational landscape when dealing with more complex systems, where achieving convergence is inherently more challenging.

\section{Conclusions}
In this work, we have presented a framework grounded on an active learning strategy for improving both accuracy and efficiency in quantum computations, with a focus on VQE. 
By employing a probabilistic machine learning based on GPR, 
our method leverages on our prior knowledge about the mathematical structure of the VQE ansatz. In particular, it accounts for exact cutoffs in the Fourier series of the variational landscape, which quantitatively encode its smoothness and periodicity properties.
A remarkable property of this framework is its compatibility with any VQE energy-minimization framework or error-mitigation method, enabling the integration of precise information about the variational landscape that is typically not utilized in standard approaches.
We applied our active learning method in combination with SPSA, a standard optimization method, to a 2-sites Anderson impurity model and to a 8-sites Heisember model, demonstrating its effectiveness across a range of parameters regimes, from weakly interacting to strongly correlated systems. Our results reveal that the SPSA+AL algorithm consistently outperforms the bare SPSA method, delivering more accurate and reliable ground-state energies, while requiring considerably fewer direct QPU energy evaluations.
While these exploratory benchmarks are based on relatively simple systems and variational ansatzes, generalizations to more complex problems are plausible, and present promising directions for future research. 
Moreover, the efficiency and accuracy of our method could be further improved by combining our active learning strategy with gradient-descent algorithms. 
This could be achieved by leveraging GPR for estimating not only the energy landscape but also its gradient. Exploring these research directions could lead to more efficient and accurate quantum computing methodologies, contributing to the development of advanced quantum matter simulation techniques.
We anticipate that our framework will prove particularly valuable within the context of QE methods, enabling calculations with large impurities to describe dynamical-correlation effects beyond the capabilities of classical impurity solvers. In particular, we foresee promising applications to QE frameworks that require computing only the ground state of a finite Anderson impurity model, such as the recently-developed "ghost Gutzwiller approximation"~\cite{Ghost-GA,ALM_g-GA} and density matrix embedding theory~\cite{DMET}.
Incorporating these QE frameworks with VQE has the potential to accelerate the advent of quantum advantage for simulations of large molecules and periodic materials, effectively capitalizing on the potential of near-term QPUs. 
In conclusion, our approach lays a solid foundation for future research and development in quantum error mitigation and optimization, potentially bringing us closer to realizing the full potential of quantum computing in the context of quantum matter simulation.

\section*{Acknowledgements}
We thank Gunnar Schmitz for useful discussions.
This work was supported by a grant from the Simons Foundation (1030691, NL). 
NL and OC gratefully acknowledge funding from the Novo Nordisk Foundation through the Exploratory Interdisciplinary Synergy Programme project NNF19OC0057790. 
OC acknowledges support from NNF20OC0065479.
Part of the quantum computing method development, implementation, calculations and analyses by YY and TJ were supported by the U.S. Department of Energy, Office of Science, National Quantum Information Science Research Centers, Co-design Center for Quantum Advantage (C2QA) under contract number DE-SC0012704. The research of YY and TJ was performed at the Ames National Laboratory, which is operated for the U.S. Department of Energy by Iowa State University under Contract No.~DE-AC02-07CH11358.
We acknowledge use of the IBM Quantum Experience, through
the IBM Quantum Researchers Program.
The views expressed are those of the authors, and do not reflect the official policy or position of IBM or the IBM Quantum team.


%

\newaliascnt{suppeqn}{equation}
\let\oldtheequation\theequation
\let\theequation\thesuppeqn

\clearpage


\onecolumngrid
\begin{center}
\textbf{\large Supplemental Information for: \\
Active Learning approach to simulations of Strongly Correlated Matter with the
Ghost Gutzwiller Approximation}
\vspace{0.4cm}
\end{center}

\setcounter{equation}{0}
\setcounter{figure}{0}
\setcounter{section}{0}
\setcounter{table}{0}
\setcounter{page}{1}
\makeatletter

\newpage

\maketitle

\section{Path integral formulation of parametric GPR}

Our goal is to learn a real valued function $\mathcal{E}(\btheta)$ ($\btheta\in\mathbb{R}^d$) from a finite set of training data points:
\be
D=\{(\btheta_\alpha,\mathcal{E}_\alpha,\sigma_\alpha)\,|\,\alpha=1,..,n\}\,, 
\label{data}
\ee
where each $\mathcal{E}_\alpha$ is the outcome of the evaluation of $\mathcal{E}$ for the input parameter $\btheta_\alpha$, which is assumed to be sampled from a gaussian distribution:
\begin{align}
P(\mathcal{E}_\alpha|\btheta_\alpha)
\propto
\exp\left\{-\frac{1}{2\sigma^2_\alpha}
(\mathcal{E}(\btheta_\alpha)-\mathcal{E}_\alpha)^2
\right\}
\,,
\end{align}
i.e., the probability of evaluating $\mathcal{E}_1,..,\mathcal{E}_n$ (which are assumed to be independent) from a given underlying function $\mathcal{E}(\btheta)$ is assumed to be:
\begin{align}
P[D|\mathcal{E}]&\propto
\prod_{\alpha=1}^n
P(\btheta_\alpha|\mathcal{E}_\alpha)
\nonumber
\\
&\propto
\exp\left\{-\sum_{\alpha=1}^n 
\frac{1}{2\sigma^2_\alpha}
(\mathcal{E}(\btheta_\alpha)-\mathcal{E}_\alpha)^2
\right\}
\label{PDf}
\,,
\end{align}

Specifically, we aim to compute the so-called "posterior probability distribution" $P[\mathcal{E}|D]$, i.e., the probability that the function that we aim to learn is $\mathcal{E}(\btheta)$, based on: (I) the data $D$ at our disposal and (II) a gaussian "prior probability distribution" $P[\mathcal{E}]$, encoding our prior knowledge before having any training data points.

Our first goal is to define precisely the concept of a probability distribution over a space of functions.
Following the path integral procedure, this can be accomplished by first considering a discrete finite mesh with uniform spacing $\epsilon$, over a $d$-dimensional rectangle $R$:
\begin{equation}
M_{\epsilon}=\{\btheta_1,..,\btheta_N\}
\subset R\subset \mathbb{R}^d
\,.
\end{equation}

Over such discretized domain, probability measures can be rigorously represented as $p_\epsilon[\mathcal{E}]\mathcal{D}_\epsilon[\mathcal{E}]$, where:
\begin{equation}
p_\epsilon[\mathcal{E}]=
p_\epsilon[\mathcal{E}(\btheta_1),..,\mathcal{E}(\btheta_N)]
\end{equation}
is a standard $N$-dimensional probability function, and:
\begin{equation}
\mathcal{D}_\epsilon[\mathcal{E}]=\prod_{\btheta\in M_{\epsilon}} d\mathcal{E}(\btheta)
\end{equation}
is the standard path integral measure.

\subsection{The parametric prior}

In our context of application, the prior probability distribution is designed to enforce the fact that $\mathcal{E}$ has to be of the following mathematical form:
\be
\mathcal{E}(\btheta)=\sum_{s=1}^{S} \xi_s T_s(\btheta)\,,
\label{parametric_form-SM}
\ee
where $T_s:R\subset\mathbb{R}^d\rightarrow \mathbb{R}$ are known functions, while 
the coefficients $\xi_s$ are unknown.
This information can be encoded in the following probability distribution:
\begin{align}
P_{\epsilon}^\eta [\mathcal{E}]&
\propto
\int \prod_{r=1}^S d\xi_r\,
e^{-\frac{\epsilon}{2\eta^2}
\sum_{\btheta\in M_{\epsilon}} 
\big(\mathcal{E}({\btheta})-\sum_s\xi_s T_s({\btheta})\big)^2
}
\nonumber
\\
&\times\,  e^{-\frac{t}{2}\epsilon
\sum_{\btheta\in M_{\epsilon}} 
\mathcal{E}^2({\btheta})}
\label{Petaepsilon}
\,,
\end{align}
where we have introduced the hyperparameter $t>0$, whose role is to make the probability distribution normalizable by enforcing that the range of $\mathcal{E}$ is bounded as we are going to prove below. 
The parameter $\eta$ will be considered in the limit as it approaches zero (i.e., we will take the limit $\eta \rightarrow 0$ later in our formalism).

Let us prove that $P_{\epsilon}^\eta [\mathcal{E}]$
is a normalizable gaussian probability distribution with zero mean for all finite values of $\eta$ and $t$.
By performing the gaussian integral in Eq.~\eqref{Petaepsilon}, we obtain that:
\begin{align}
    P_{\epsilon}^\eta [\mathcal{E}]
    \propto
    e^{-\frac{1}{2}
\sum_{\btheta,\btheta'\in M_{\epsilon}} 
\epsilon\left[
t\mathds{1}+\frac{1}{\eta^2}\Pi
\right]_{\btheta\btheta'}
\mathcal{E}({\btheta})\mathcal{E}({\btheta}')
}
\label{Petaepsilon-matrix}
\,,
\end{align}
where $\mathds{1}$ is the $N\times N$ identity matrix and:
\begin{align}
    \Pi&=\mathds{1}-T(T^\dagger T)^{-1}T^\dagger
    \\
    T_{\btheta s}&=T_s(\btheta)\,\forall\,s=1,..,S,\,\btheta\in
    M_\epsilon
    \,.
\end{align}
Note that $\Pi$ is an orthogonal projector and, therefore, it is positive semi-definite.
It follows that Eq.~\eqref{Petaepsilon-matrix} represents a normalizable zero-mean gaussian distribution $\forall$ $t>0$.

\subsection{Posterior probability distribution}

Let us assume to have a series of data $D$ (see Eq.~\eqref{data}), where 
$\btheta_\alpha\in M_\epsilon$ $\forall\,\alpha=1,..,n$.
As explained in the main text (see before Eq.20),
from Bayes' theorem it follows that the posterior
conditional probability distribution for the function $\mathcal{E}$ is the following:
\be
P^\eta_\epsilon[\mathcal{E}|D]\propto P^\eta_\epsilon[\mathcal{E}] 
e^{-\sum_{\alpha=1}^n 
\frac{1}{2\sigma^2_\alpha}
(\mathcal{E}(\btheta_\alpha)-\mathcal{E}_\alpha)^2
}
\,,
\label{posterior}
\ee
which represents the probability distribution for the function $\mathcal{E}$, given the data set $D$ and the prior $P^\eta_\epsilon[\mathcal{E}]$ (see Eqs.~\eqref{PDf} and \eqref{Petaepsilon}).

\subsection{Probabilistic predictions at a test point}

We are interested in calculating quantities of the following form:
\begin{align}
    \langle \mathcal{E}^l(\btheta)\rangle
    &=
    \int \mathcal{D}_\epsilon[\mathcal{E}]
    P^\eta_\epsilon[\mathcal{E}|D]\,\left(\mathcal{E}(\btheta)\right)^l
    \label{integral-posterior}
    \\
    &=\int \mathcal{D}_\epsilon[\mathcal{E}]
    P^\eta_\epsilon[\mathcal{E}]\,
    e^{-\sum_{\alpha=1}^n 
\frac{1}{2\sigma^2_\alpha}
(\mathcal{E}(\btheta_\alpha)-\mathcal{E}_\alpha)^2
}\mathcal{E}^l(\btheta)
    \nonumber
    \,,
\end{align}
where $l\in\mathbb{N}$ and $\btheta\in M_\epsilon$ (which is assumed to be different from all of the $\btheta_\alpha$ in the training data set) is a so-called "test point," i.e., a point where we want to evaluate the probability distribution for $\mathcal{E}(\btheta)$, based on our posterior probability distribution.

Eq.~\eqref{integral-posterior} can be conveniently rewritten by integrating out all variables except $\mathcal{E}(\btheta_1),..,\mathcal{E}(\btheta_\alpha)$ and $\mathcal{E}(\btheta)$. 
From standard Gaussian identities, it follows that this gives the following expression:
\begin{align}
    \langle \mathcal{E}^l(\btheta)\rangle
    &=
    \frac{
    \int \big[\prod_{\alpha=1}^n
    d\mathcal{E}(\btheta_\alpha)\big] d\mathcal{E}(\btheta)
    \,
    e^{-S^\eta_\epsilon-U}
    \,\left(\mathcal{E}(\btheta)\right)^l
    }
    {
\int \big[\prod_{\alpha=1}^n
    d\mathcal{E}(\btheta_\alpha)\big] d\mathcal{E}(\btheta)
    \,
    e^{-S^\eta_\epsilon-U}
    }
    \,,
\end{align}
where:
\begin{align}
    U &=\sum_{\alpha=1}^n 
\frac{1}{2\sigma^2_\alpha}
(\mathcal{E}(\btheta_\alpha)-\mathcal{E}_\alpha)^2
\end{align}
and:
\begin{align}
    S^\eta_\epsilon &=
    \frac{1}{2}\sum_{\alpha,\beta=1}^n
    [\bar{\mathcal{K}}_\epsilon^\eta]^{-1}_{\alpha,\beta}
    \,
    \mathcal{E}(\btheta_\alpha)\mathcal{E}(\btheta_\beta)
    \nonumber\\
    &+
    \frac{1}{2}
    [\bar{\mathcal{K}}_\epsilon^\eta]^{-1}_{n+1,n+1}
    \,
    \mathcal{E}(\btheta)\mathcal{E}(\btheta)
    \nonumber\\
    &+
    \frac{1}{2}\sum_{\alpha=1}^n
    [\bar{\mathcal{K}}_\epsilon^\eta]^{-1}_{\alpha,n+1}
    \,
    \mathcal{E}(\btheta_\alpha)\mathcal{E}(\btheta)
    \nonumber\\
    &+
    \frac{1}{2}\sum_{\beta=1}^n
    [\bar{\mathcal{K}}_\epsilon^\eta]^{-1}_{n+1,\beta}
    \,
    \mathcal{E}(\btheta)\mathcal{E}(\btheta_\beta)
    \,,
\end{align}
where $[\bar{\mathcal{K}}_\epsilon^\eta]$ is the $(n+1)\times (n+1)$ matrix with entries:
\begin{align}
[\bar{\mathcal{K}}_\epsilon^\eta]_{\alpha,\beta}&=K_\epsilon^\eta(\btheta_\alpha,\btheta_\beta)
\quad
\forall\,\alpha,\beta\in 1,..,n
\\
[\bar{\mathcal{K}}_\epsilon^\eta]_{\alpha,n+1}&=K_\epsilon^\eta(\btheta_\alpha,\btheta)
\quad
\forall\,\alpha\in 1,..,n
\\
[\bar{\mathcal{K}}_\epsilon^\eta]_{n+1,\beta}&=K_\epsilon^\eta(\btheta,\btheta_\beta)
\quad
\forall\,\beta\in 1,..,n
\\
[\bar{\mathcal{K}}_\epsilon^\eta]_{n+1,n+1}&=K_\epsilon^\eta(\btheta,\btheta)
\end{align}
and
\begin{align}
K_\epsilon^\eta(\btheta,\btheta')=\int \D_\epsilon[\mathcal{E}] P_\epsilon^\eta[\mathcal{E}]\, \mathcal{E}(\btheta)\,\mathcal{E}(\btheta')
\;\;\forall\,\btheta,\btheta'\in M_\epsilon
\label{kernel-definition}
\end{align}
is the so-called "kernel function" of the prior distribution $P_\epsilon^\eta$.

As discussed in the main text, we are specifically interested in calculating:
\begin{align}
\bar{\mathcal{E}}_\epsilon^\eta(\btheta) &= \int \D_\epsilon[\mathcal{E}] P_\epsilon^\eta[\mathcal{E}|D]\, \mathcal{E}(\btheta)
\label{av-s}
\\
\left(\Sigma_\epsilon^\eta(\btheta)\right)^2  &= \int \D_\epsilon[\mathcal{E}] P_\epsilon^\eta[\mathcal{E}|D]\,
\big(\mathcal{E}^2(\btheta)-\langle \mathcal{E}(\btheta) \rangle^2\big)
\label{sq-s}
\,,
\end{align}
where Eq.~\eqref{av-s} represents our prediction for $\mathcal{E}(\btheta)$ at any test point $\btheta$ and Eq.~\eqref{sq-s} represents the uncertainty of our prediction.
These quantities can be conveniently evaluated by computing first the "partition function":
\begin{equation}
    Z_\epsilon^\eta(\lambda):=
    \int \big[\prod_{\alpha=1}^n
    d\mathcal{E}(\btheta_\alpha)\big] d\mathcal{E}(\btheta)
    \,
    e^{-S^\eta_\epsilon-U+\lambda\mathcal{E}(\btheta)
    }
\end{equation}
and subsequently using the following identities:
\begin{align}
\bar{\mathcal{E}}_\epsilon^\eta(\btheta) &= 
\partial_\lambda \ln(Z_\epsilon^\eta(\lambda))
\label{av-sp}
\\
\left(\Sigma_\epsilon^\eta(\btheta)\right)^2  &= 
\partial^2_\lambda \ln(Z_\epsilon^\eta(\lambda))
\label{sq-sp}
\,.
\end{align}
A direct calculation shows that:
\begin{align}
\bar{\mathcal{E}}_\epsilon^\eta(\btheta) &= 
\sum_{\alpha,\beta=1}^n
K_\epsilon^\eta(\btheta,\btheta_\alpha)
[\bar{K}_\epsilon^\eta]^{-1}_{\alpha\beta}\mathcal{E}_\beta
\label{av-explicit-s}
\\
\left(\Sigma_\epsilon^\eta(\btheta)\right)^2 &= 
K_\epsilon^\eta(\btheta,\btheta)-\!\!
\sum_{\alpha,\beta=1}^n \!\!
K_\epsilon^\eta(\btheta,\btheta_\alpha)
[\bar{K}_\epsilon^\eta]^{-1}_{\alpha\beta}
K_\epsilon^\eta(\btheta_\beta,\btheta)
\,,
\label{sq-explicit-s}
\end{align}
where $\bar{K}_\epsilon^\eta$ is the $n\times n$ matrix with entries:
\begin{align}
[\bar{K}_\epsilon^\eta]_{\alpha\beta}=K_\epsilon^\eta(\btheta_\alpha,\btheta_\beta)+
\sigma^2_\alpha\delta_{\alpha\beta}\quad
\forall\,\alpha,\beta\in 1,..,n
\,.
\end{align}

\subsection{Calculation of the Kernel function}

As shown in the previous section, the GPR estimate of our prediction for $\mathcal{E}(\btheta)$ and the corresponding uncertainty (see Eqs.~\eqref{av-explicit-s} and \eqref{sq-explicit-s}, respectively) depend explicitly on $\eta$
and $\epsilon$ through the Kernel function:
\begin{align}
K_\epsilon^\eta(\btheta_1,\btheta_2)=\int \D_\epsilon[\mathcal{E}] P_\epsilon^\eta[\mathcal{E}] \, \mathcal{E}(\btheta_1)\mathcal{E}(\btheta_2)
\,,
\label{kernel-definition-2}
\end{align}
which is defined $\forall\,\btheta_1,\btheta_2\in M_\epsilon$.

Since we aim to enforce Eq.~\eqref{parametric_form-SM} \emph{exactly}, we need to evaluate Eq.~\eqref{kernel-definition-2} for $\eta\rightarrow 0$.
In this limit we obtain:
\begin{widetext}
\begin{align}
    K_\epsilon(\btheta_1,\btheta_2)
    &=\lim_{\eta\rightarrow 0}
    K_\epsilon^\eta(\btheta_1,\btheta_2)
    \nonumber
    \\
    &\propto\lim_{\eta\rightarrow 0}
    \int \D_\epsilon[\mathcal{E}] 
    \int \prod_{r=1}^S d\xi_r\,
e^{-\frac{\epsilon}{2\eta^2}
\sum_{\btheta\in M_{\epsilon}} 
\big(\mathcal{E}({\btheta})-\sum_s\xi_s T_s({\btheta})\big)^2
    }
    e^{-\frac{t}{2}\epsilon
\sum_{\btheta\in M_{\epsilon}} 
\mathcal{E}^2({\btheta})}
    \, \mathcal{E}(\btheta_1)\,\mathcal{E}(\btheta_2)
    \nonumber
    \\
    &\propto
    \int \D_\epsilon[\mathcal{E}] 
    \int \prod_{r=1}^S d\xi_r\,
\delta
\big(\mathcal{E}({\btheta})-\sum_s\xi_s T_s({\btheta})\big)
\,
e^{-\frac{t}{2}\epsilon
\sum_{\btheta\in M_{\epsilon}} 
\mathcal{E}^2({\btheta})}
    \, \mathcal{E}(\btheta_1)\,\mathcal{E}(\btheta_2)
     \nonumber
    \\
    &=
    \int \prod_{r=1}^S d\xi_r\,
    e^{-\frac{t}{2}\epsilon
\sum_{\btheta\in M_{\epsilon}} 
\left(\sum_{s=1}^S\xi_s T_s(\btheta)\right)^2}
\,
\left(\sum_{s_1=1}^S\xi_{s_1} T_{s_1}(\btheta_1)\right)
\left(\sum_{s_2=1}^S\xi_{s_2} T_{s_2}(\btheta_2)\right)
     \nonumber
    \\
    &=
    \sum_{s_1,s_2=1}^ST_{s_1}(\btheta_1)T_{s_2}(\btheta_2)
    \int \prod_{r=1}^S d\xi_r\,
    e^{-\frac{t}{2}
    \sum_{s,s'=1}^S \mathcal{A}^\epsilon_{ss'}\,\xi_s\xi_{s'}
\label{calculation-K}
}
\,
\xi_{s_1} \xi_{s_2} 
     \nonumber
    \\
    &=
    t^{-1}\sum_{s_1,s_2=1}^S \Delta^\epsilon_{s_1s_2}
    \,
    T_{s_1}(\btheta_1)T_{s_2}(\btheta_2)
    \,,
\end{align}

\end{widetext}
where:
\begin{align}
\Delta^\epsilon &=[\mathcal{A}^\epsilon]^{-1}
\\
\mathcal{A}^\epsilon_{ss'}&=
    \epsilon
    \sum_{\btheta\in M_{\epsilon}} 
T_s(\btheta)T_{s'}(\btheta)
\,.
\end{align}

The final step is to compute the Kernel function in the continuum limit $\epsilon\rightarrow 0$, which is given by the following equation:
\begin{align}
    K(\btheta_1,\btheta_2)&=\lim_{\epsilon\rightarrow 0}K_\epsilon(\btheta_1,\btheta_2)
    \nonumber\\
    &=t^{-1}\sum_{s_1,s_2=1}^S \Delta_{s_1s_2}\,
    T_{s_1}(\btheta)T_{s_2}(\btheta)
    \,,
\end{align}
where:
\begin{align}
\Delta &=\mathcal{A}^{-1}
\\
\mathcal{A}_{ss'}&=\lim_{\epsilon\rightarrow 0}\mathcal{A}^\epsilon_{ss'}
=\lim_{\epsilon\rightarrow 0}\,
    \epsilon\sum_{\btheta\in M_{\epsilon}} 
T_s(\btheta)T_{s'}(\btheta)
\nonumber\\
&=\int_R d\btheta \, T_s(\btheta)T_{s'}(\btheta)
\,.
\end{align}

Note that the calculation of $K(\btheta_1,\btheta_2)$
in Eq.~\eqref{calculation-K}
becomes straightforward if the the functions 
$T_s(\btheta)$ are replaced by an orthonormal basis $\tau_\bk(\btheta)$ of the same space with respect to the 
$L^2(R)$ metric, as in the main text.
In fact, with such a choice we obtain that:
\begin{align}
    K(\btheta_1,\btheta_2)&=t^{-1}\sum_{\bk=1}^S\,
    \tau_{\bk}(\btheta)\tau_{\bk}(\btheta)
    \,,
    \label{kernel-orthonormal}
\end{align}
which is practically more convenient because:
(i) there is no need to invert the matrix $\A$ (which may become
prohibitive for high-dimensional spaces), and (ii) 
evaluating Eq.~\eqref{kernel-orthonormal} involves a single summation rather than a double summation, which makes it
less computationally demandint to evaluate.

\subsection{Summary of final equations}

In summary, by replacing the kernel function $K_\epsilon^\eta(\btheta,\btheta')$ with:
\begin{equation}
K(\btheta,\btheta')
=\lim_{\epsilon\rightarrow 0}
\lim_{\eta\rightarrow 0}K_\epsilon^\eta(\btheta,\btheta')
\,,
\end{equation}
we obtain the equations shown in the main text, i.e.:
\begin{align}
\bar{\mathcal{E}}(\btheta) &= 
\sum_{\alpha,\beta=1}^n
K(\btheta,\btheta_\alpha)
[\bar{K}]^{-1}_{\alpha\beta}\mathcal{E}_\beta
\label{av-explicit-s}
\\
\left(\Sigma(\btheta)\right)^2 &= 
K(\btheta,\btheta)-\!\!
\sum_{\alpha,\beta=1}^n \!\!
K(\btheta,\btheta_\alpha)
\bar{K}^{-1}_{\alpha\beta}
K(\btheta_\beta,\btheta)
\,,
\label{sq-explicit-s}
\end{align}
where $\bar{K}$ is the $n\times n$ matrix with entries:
\begin{align}
[\bar{K}]_{\alpha\beta}=K(\btheta_\alpha,\btheta_\beta)+
\sigma^2_\alpha\delta_{\alpha\beta}\quad
\forall\,\alpha,\beta\in 1,..,n
\,.
\end{align}


\end{document}